\documentclass[conference,a4paper]{IEEEtran}

\usepackage{multirow}% http://ctan.org/pkg/multirow
\usepackage[font=scriptsize,caption=false,labelsep=space]{subfig}
\usepackage{mathrsfs}
\usepackage{graphicx,float,wrapfig,epstopdf,amsmath}
\usepackage[]{algorithmicx}
\usepackage{algpseudocode,algorithm}
\usepackage{balance}
\usepackage{enumitem}
\usepackage{cuted}

\usepackage{mathtools,lipsum}
\usepackage{amsmath}
\usepackage{amssymb}
\usepackage{amsthm}
\usepackage{graphicx}
\usepackage{epstopdf}
\usepackage{times}
\usepackage{textcomp,cite}

\usepackage{url}
\usepackage{hyperref}

\usepackage{multirow}
\usepackage{threeparttable}
\graphicspath{{./figures/}}

\usepackage[table]{xcolor}% http://ctan.org/pkg/xcolor
\definecolor{infocolor}{RGB}{213,229,255}
\definecolor{inteins}{RGB}{128,179,255}
\definecolor{color1}{RGB}{199,209,232}
\definecolor{color2}{RGB}{230,231,233}

\usepackage[table]{xcolor}% http://ctan.org/pkg/xcolor
\newtheorem{theorem}{Theorem}

\newtheorem{proposition}[theorem]{Proposition}

\usepackage{setspace}

\begin{document}

	%TC:ignore
	\title{Spherical Wavefront Near-Field DoA Estimation in THz Automotive Radar   }

	\author{
		\IEEEauthorblockN{Ahmet~M. Elbir$^{1}$, Kumar Vijay Mishra$^{1,2}$ and Symeon Chatzinotas$^1$	}
		\IEEEauthorblockA{$^{1}$University of Luxembourg, Luxembourg\\ 
			$^2$United States DEVCOM Army Research Laboratory, Adelphi, MD 20783 USA\\
			E-mail: \texttt{ahmetmelbir@ieee.org, kvm@ieee.org, symeon.chatzinotas@uni.lu}
			%			\vspace{-8pt}
		}
	}
	
	\maketitle
	\begin{abstract}
		Automotive radar at terahertz (THz) band has the potential to provide compact design. The availability of wide bandwidth at THz-band leads to high range resolution. Further, very narrow beamwidth arising from large arrays yields high angular resolution up to milli-degree level direction-of-arrival (DoA) estimation. At THz frequencies and extremely large arrays, the signal wavefront is spherical in the near-field that renders traditonal far-field DoA estimation techniques unusable. In this work, we examine near-field DoA estimation for THz automotive radar. We propose an algorithm using multiple signal classification (MUSIC) to estimate target DoAs and ranges while also taking beam-squint in near-field into account. Using an array transformation approach, we compensate for near-field beam-squint in noise subspace computations to construct the beam-squint-free MUSIC spectra. Numerical experiments show the effectiveness of the proposed method to accurately estimate the target parameters. 		
	\end{abstract}

	\vskip0.5\baselineskip
	%
	
	%TC:endignore
	
	%	\setstretch{1}
	%	
	
	%	\vspace{-12pt}
	\section{Introduction}
	\label{sec:Introduciton}
	\IEEEPARstart{T}{erahertz} (THz) band, spanning from $0.1$ to $10$ THz, has emerged as a promising frontier for the realization of significant advancements in sixth-generation (6G) wireless networks~\cite{thz_Akyildiz2022May}. Apart from gains in the communications performance, THz-band is also currently invetsigated to ensure milli-degree precision in direction-of-arrival (DoA) estimation in THz automotive radar~\cite{milliDegree_doa_THz_Chen2021Aug,milliDegreeDOA_THz_Peng2016Aug,elbir2022Aug_THz_ISAC}. Other advantages of THz-band sensing include small form factor, wide bandwidth, and near-optical resolution~\cite{thzRadar1_Bhattacharjee,thzRadar2_Xiao2020Mar,elbir2021JointRadarComm,eamaz2023near,mishra2023signal}.  
	%	Currently, low-THz 	frequencies, such as 0.15 THz and 0.3 THz bands, provide 6 GHz 	and 16 GHz unlicensed contiguous bandwidths, respectively. These 	wide frequency bands enable automotive radars to achieve lidar-like 	imaging capabilities~\cite{}. 

	%	Furthermore, low-THz spectrum exhibits distance-dependent spectral windows, whose bandwidth shrinks with the transmission distance, especially when this distance is increased from 1 to 10 meters due to high molecular absorption~\cite{elbir2022Aug_THz_ISAC}. As a result, THz-band implementation exhibits a trade-off between operating the automotive radar at large bandwidth to improve range resolution and maintaining a maximum detectable range. 

\begin{figure}[t]
		\centering
		{\includegraphics[draft=false,width=\columnwidth]{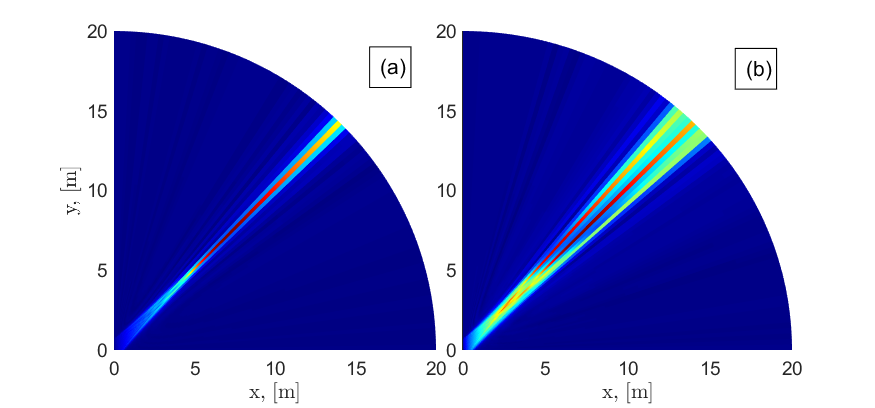} } 
		\caption{ Array gain of a single target at $45^\circ$ and $10$ m for (a) narrowband  and (b) wideband scenario when $f_c=300$ GHz, $B=10$ MHz and $B = 30$ GHz, respectively.
		}
		%					\vspace*{-5mm}
		\label{fig_ArrayGain}
	\end{figure}

	High-resolution DoA estimation within the THz-band, however, is impeded by myriad challenges such as high path losses, molecular absorption, and intricate propagation/scattering dynamics \cite{ummimoTareqOverview,thz_Akyildiz2022May}. To compensate these losses, large number of antennas are employed to improve the beamforming gain~\cite{elbir2022Nov_SPM_beamforming}. Furthermore, hybrid analog/digital beamforming architectures with phase shifter networks are used to reduce the number of radio-frequency (RF) chains. THz systems also suffer from \textit{beam-squint} arising from the subcarrier-independent analog beamformers~\cite{delayPhasePrecoding_THz_Dai2022Mar,beamSquint_FeiFei_Wang2019Oct,elbir_THZ_CE_ArrayPerturbation_Elbir2022Aug}. This leads to misaligned beam generation at different subcarrier-squint %and point to different directions
	in the spatial domain; that is, the main lobes of the array gain corresponding to the lowest and highest subcarriers do not overlap because of ultra-wide bandwidth (Fig.~\ref{fig_ArrayGain}). This significantly degrades DoA estimation~\cite{elbir2022Aug_THz_ISAC,elbir_THZ_CE_ArrayPerturbation_Elbir2022Aug}.  
	
	Notably, existing countermeasures for beam-squint are predominantly hardware-based~\cite{thz_beamSplit}. Here, additional hardware components such as time-delayer networks are realized to generate a negative group-delay for its compensation~\cite{beamSquint_FeiFei_Wang2019Oct}. This approach is expensive because each phase shifter of the network is connected to multiple delayer elements, each of which consumes approximately $150\%$  more power than a single phase shifter at THz band~\cite{elbir2022Aug_THz_ISAC}. THz channel estimation~\cite{elbir_THZ_CE_ArrayPerturbation_Elbir2022Aug} and hybrid analog/digital beamforming~\cite{delayPhasePrecoding_THz_Dai2022Mar,beamSquint_FeiFei_Wang2019Oct,elbir2021JointRadarComm} under beam-squint have been explored in prior THz studies, which largely omit discussions on DoA estimation.	While the DoA estimation problem is studied for both THz~\cite{milliDegree_doa_THz_Chen2021Aug} and millimeter-wave~\cite{calibrationMassiveMIMOWei2020Apr,doaEst_mmWave_Zhang2021Oct}, the impact of beam-squint is generally excluded from such studies.
	
	Besides beam-squint, another formidable challenge in THz-band signal processing is short-transmission distance. While some experiments indicate that a maximum range of $200$ m is possible for THz automotive radars  \cite{norouzian2019rain}, practical operations are envisaged only in the 10-20 m range\cite{thzRadar1_Bhattacharjee}. Together with high THz frequencies and usage of extremely large (XL) arrays, the received signal wavefront at close ranges is spherical in near-field (Fig.~\ref{fig_NF}). In general, the wavefront is spherical in the near-field when the transmission range is shorter than the Fraunhofer distance~\cite{nf_primer_Bjornson2021Oct,elbir2023Feb_NF_THZ_CE_ICASSP_NBAOMP}. THz automotive radars must, therefore, accommodate the near-field beampattern for DoA estimation, which now depends on both direction and range information~\cite{elbir2022Aug_THz_ISAC}. Among prior works, \cite{nf_OMP_Dai_Wei2021Nov,nf_mmwave_CE_noBeamSplit_Cui2022Jan,nf_NB2_Zhang2022Nov,eamaz2023near} consider near-field processing for THz systems but ignore the effect of beam-squint. % and focusing solely on narrowband millimeter-wave scenarios.
	Range-dependent beampattern is also observed in some far-field applications such as frequency diverse array (FDA) radars \cite{lv2022co,lv2022clutter} and quantum Rydberg arrays \cite{vouras2023overview,vouras2023phase}. However, the wavefront is not spherical in these applications.
	
	In this paper, we examine the near-field DoA estimation problem for THz automotive radar in the presence of beam-squint. We first present the near-field signal model as well as the beam-squint model. Then, a subarrayed approach is devised to collect and process the antenna array outputs for high resolution DoA estimation. We propose a modified \textit{mu}ltiple \textit{si}gnal \textit{c}lassification (MUSIC) algorithm~\cite{music_Schmidt1986Mar} with noise  subspace correction approach  to account for the beam-squint and accurately estimate the targets. In order to obtain the corrected noise subspace, we introduce a linear transformation matrix that constructs a mapping the nominal and the beam-squint-distorted steering vectors in spherical wave domain, which facilitates the rectification of the skewed noise-subspace matrix derived from the covariance of the array data. The performance of the proposed approach is evaluated via numerical experiments, and it can effectively estimate the target DoA angles and ranges in the presence of near-field beam-squint.

	%	We also derive the Cram\'er-Rao lower bound (CRB) for the proposed model to benchmark our DoA estimates. %, which is derived for the considered problem.
	
	%%-----------------------------------------------------
	%%-----------------------------------------------------
	
	\section{System Model}
	\label{sec:SignalModel}
\begin{figure}[t]
		\centering
		{\includegraphics[draft=false,width=\columnwidth]{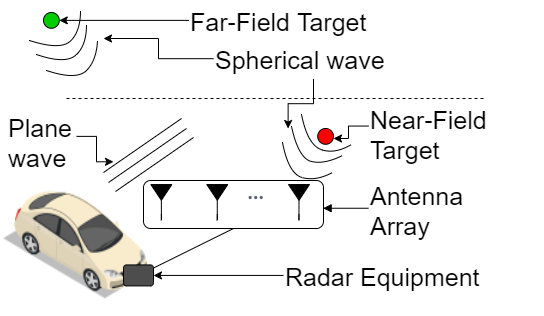} } 
		\caption{An illustration of THz automotive radar and the received signal wavefronts from far- and near-field targets. 
		}
		%					\vspace*{-5mm}
		\label{fig_NF}
	\end{figure}

	Consider a wideband THz automotive radar system (Fig.~\ref{fig_NF}) that comprises hybrid analog/digital beamformers performed over $M$ subcarriers with $N$-element uniform linear array (ULA) and $N_\mathrm{RF}$ RF chains. The radar employs a subcarrier-independent precoder $\mathbf{F}\in \mathbb{C}^{N\times N_\mathrm{RF}}$ to sense the environment. The sensing signal generated for the $m$-th subcarrier is  $\mathbf{s}_m(t_i)\in \mathbb{C}^{N_\mathrm{RF}}$, where $i = 1,\cdots, T$,  and $T$ is the number of snapshots along the fast-time axis~\cite{mimoRadar_WidebandYu2019May}. To sense the environment,  $N_\mathrm{RF}$ RF chains are activated and the transmitted probing signal is 
	\begin{align}
	\mathbf{X}_m = \left[\mathbf{x}_m(t_1),\cdots, \mathbf{x}_m(t_T) \right]\in \mathbb{C}^{N\times T},
	\end{align}
	where $\mathbf{x}_m(t_i) = \mathbf{F}\mathbf{s}_m(t_i)\in \mathbb{C}^{N}$,  $\mathbb{E}\{ \mathbf{X}_m  \mathbf{X}_m^\textsf{H} \} =\frac{P_rT}{MN} \mathbf{I}_N$, $\mathbf{FF}^\textsf{H} = 1/N$, and $P_r$ is the radar transmit power.	 
	
	\subsection{Near-field}
	The use of high THz frequencies and XL arrays implies that the short-range targets may lie in the near-field region, wherein the planar wave propagation does not hold. At ranges shorter than the Fraunhofer distance $d_F = \frac{2 D^2}{\lambda}$, where $D$ is the array aperture and $\lambda = \frac{c_0}{f_c}$ is the wavelength, the near-field wavefront is spherical  \cite{nf_primer_Bjornson2021Oct,elbir_THZ_CE_ArrayPerturbation_Elbir2022Aug}.  For a ULA, the array aperture is $D = (N-1)d$, where $d = \frac{\lambda}{2}$ is the element spacing. In the THz spectrum, it is imperative to employ a near-field signal model because $r_{k} <d_F$. For instance, when $f_c = 300$ GHz and $N=256$, the Fraunhofer distance is $d_F = 32.76$ m.
	
	Suppose that there are $K$ targets at the physical locations $\{\theta_k, r_k \}_{k = 1}^K$, where $\theta_{k}$ and $r_k$ denote the DoA and the range of the $k$-th target. Taking into account the spherical-wave model~\cite{nf_primer_Bjornson2021Oct,nf_Fresnel_Cui2022Nov,elbir2023Feb_NF_THZ_CE}, define the near-field steering vector $\mathbf{a}(\theta_{k},r_{k})\in\mathbb{C}^{N}$ corresponding to the physical DoA  $\theta_{k}$ and range $r_{k}$ as 
	\begin{align}
	\label{steeringVec1}
	\mathbf{a}(\theta_{k},r_{k}) = \frac{1}{\sqrt{N}} [e^{- \mathrm{j}2\pi \frac{d}{\lambda}r_{k}^{(1)} },\cdots,e^{- \mathrm{j}2\pi \frac{d}{\lambda}r_{k}^{(N)} }]^\textsf{T},
	\end{align}
	where  $r_{k}^{(n)}$ is the distance between the $k$-th target and the $n$-th radar antenna, i.e.,
	\begin{align}
	r_{k}^{(n)} = \sqrt{r_{k}^2  + 2(n-1)^2 d^2 - 2 r_{k}(n-1) d \theta_{k}   }. \label{eq:rkln}
	\end{align}
	Following the Fresnel approximation~\cite{nf_Fresnel_Cui2022Nov,elbir2023Feb_NF_THZ_CE}, \eqref{eq:rkln} becomes
	\begin{align}
	\label{r_approx}
	r_{k}^{(n)} \approx r_{k}  - (n-1) d \theta_{k}  + (n-1)^2 d^2 \zeta_{k}  ,
	\end{align}	 
	where $\zeta_{k} = \frac{1- \theta_{k}^2}{2 r_{k}}$. Rewrite (\ref{steeringVec1}) as
	\begin{align}
	\label{steeringVectorPhy}
	\mathbf{a}(\theta_{k},r_{k}) \approx e^{- \mathrm{j}2\pi \frac{f_c}{c_0}r_{k}} \tilde{\mathbf{a}}(\theta_{k},r_{k}),
	\end{align} where the $n$-th element of $\tilde{\mathbf{a}}(\theta_{k},r_{k})\in \mathbb{C}^{N}$ is 
	\begin{align}
	\label{steeringVectorPhy2}
	[\tilde{\mathbf{a}}(\theta_{k},r_{k})]_n = e^{\mathrm{j} 2\pi \frac{f_c}{c_0}\left( (n-1)d\theta_{k}  - (n-1)^2 d^2 \zeta_{k}\right) }.
	\end{align}
	
	\subsection{Beam-squint}
	The steering vector in (\ref{steeringVectorPhy}) corresponds to the physical location $(\theta_{k},r_{k})$. Due to beam-squint, this deviates to the spatial location $(\bar{\theta}_{m,k},\bar{r}_{m,k})$ in the beamspace because of the absence of subcarrier-dependent analog beamformers. Then, the $n$-th entry of the deviated steering vector in (\ref{steeringVectorPhy2}) for the spatial location is 
	\begin{align}
	\label{steeringVectorSpa}
	&[\tilde{\mathbf{a}}(\bar{\theta}_{m,k},\bar{r}_{m,k})]_n \hspace{-3pt}= \hspace{-2pt}e^{\mathrm{j} 2\pi \frac{f_m}{c_0}\left( (n-1)d\bar{\theta}_{m,k}  - (n-1)^2 d^2 \bar{\zeta}_{m,k}\right) }.
	\end{align}

	%%-----------------------------------------------------
	%%-----------------------------------------------------
	
	%	 We introduce the following Theorem 1 to establish the relationship between the physical and spatial DoAs/ranges. 
	Recall the following useful result to establish the relationship between the physical and spatial DoAs/ranges.

	%	 o\textcolor{red}{that ....write about this result. Since it has been taking from abother reference, no need to include proof. Just cite it. Also, make it a proposition because it is not a Theorem of this paper.}
	
	\begin{proposition}\cite{elbir2023Feb_NF_THZ_CE}
		Denote $\mathbf{u}\in \mathbb{C}^{N} $ and $\mathbf{v}_m \in \mathbb{C}^{N}$ as the  near-field steering vectors corresponding to the physical (i.e., $\{\theta_{k},r_{k}\}$) and spatial (i.e., $\{\bar{\theta}_{m,k},	\bar{r}_{m,k}\}$) locations given in (\ref{steeringVectorPhy2}) and (\ref{steeringVectorSpa}), respectively. Then, in spatial domain at subcarrier frequency $f_m$, the array gain achieved by $\mathbf{u}^\textsf{H}\mathbf{v}_m$ is maximized and the generated beam is focused at the location $\{\bar{\theta}_{m,k},	\bar{r}_{m,k}\}$ such that  
		\begin{align}
		\label{physical_spatial_directions}
		\bar{\theta}_{m,k} =    \eta_m \theta_{k}, \hspace{5pt}
		\bar{r}_{m,k} =    \frac{1 - \eta_m^2 \theta_{k}^2}{\eta_m(1 -\theta_{k}^2)}r_{k},
		\end{align}
		where  	 $\eta_m = \frac{f_c}{f_m}$ represents the proportional deviation of DoA/ranges.
	\end{proposition}
	
	%\begin{proof}
	%	See~\cite{elbir2023Feb_NF_THZ_CE}.
	%\end{proof}

	%	\vspace{-8pt}
	
	Following (\ref{r_approx}) and (\ref{physical_spatial_directions}), define the near-field beam-squint in terms of DoAs and ranges as, respectively,
	\begin{align}
	\label{beamSplit2}
	\Delta(\theta_{k},m) &= \bar{\theta}_{m,k} - \theta_{k} = (\eta_m -1)\theta_{k}, 
	\end{align}
	and 
	\begin{align}
	\Delta(r_{k},m) &= \bar{r}_{m,k} - r_{k} = (\eta_m -1)r_{k}\nonumber\\
	&=  (\eta_m -1) \frac{1 - \eta_m^2 \theta_{k}^2}{\eta_m(1 -\theta_{k}^2)}r_{k}.
	\end{align} 
	
	Define the received $N\times T$ target echo signal impinging on the antenna array at the $m$-th subcarrier  as
	\begin{align}
	\widetilde{\mathbf{Y}}_m = \sum_{k = 1}^{K}  \mathbf{a}(\bar{\theta}_{m,k},\bar{r}_{m,k}) \widetilde{\mathbf{x}}_{m,k} + \widetilde{\mathbf{N}}_m, \label{receivedSignal1}
	\end{align}
	where  	$\widetilde{\mathbf{N}}_m \sim \mathcal{CN}(\mathbf{0},\sigma^2  \mathbf{I}_{N})
	$ is temporarily and spatially white zero-mean complex Gaussian noise matrix of size $N\times T$ with variance $\sigma^2$;     $\widetilde{\mathbf{x}}_{m,k}\in\mathbb{C}^{1\times T}$ denotes the echo signal reflected from the $k$-th target as $\widetilde{\mathbf{x}}_{m,k} = \beta_{m,k} \mathbf{a}^\textsf{T}(\bar{\theta}_{m,k},\bar{r}_{m,k}) \mathbf{X}_m$; and $\beta_{m,k}\in \mathbb{C}$ is the reflection coefficient.
	
	Our goal is to estimate the DoA and ranges of the targets, given that the array outputs from $N$ antennas. 
	
	%	\textcolor{red}{Our goal is .....}
	
			\vspace*{-1mm}
	\section{Proposed Approach}
	As $N_\mathrm{RF}< N$, the size of the $N\times 1$ received signal in (\ref{receivedSignal1}) is reduced to $N_\mathrm{RF}\times 1$ after combining process as
	\begin{align}
	\label{radarReceived1}
	\breve{\mathbf{Y}}_{m}  = \sum_{k = 1}^{K} \breve{\mathbf{W}}^\textsf{H} {\mathbf{a}}(\bar{\theta}_{m,k},\bar{r}_{m,k}) \widetilde{\mathbf{x}}_{k,m} + \breve{\mathbf{W}}^\textsf{H}\widetilde{\mathbf{N}}_m,
	\end{align}
	where $\breve{\mathbf{W}}\in\mathbb{C}^{N\times N_\mathrm{RF}}$ represents the analog combiner matrix.

	%	
	%	In order to collect the full array data from $N_\mathrm{RF}$ RF chains, we follow a subarrayed approach, wherein the BS activates the antennas in a subarrayed fashion to obtain $N\times 1$ array data in $J=\frac{N}{N_\mathrm{RF}}$ time slots. Let $\mathbf{W}_j \in \mathbb{C}^{N\times N_\mathrm{RF}}$ be the applied combiner matrix at the $j$-th time slot (instead of $\breve{\mathbf{W}}$ in (\ref{radarReceived1})) as $\mathbf{W}_j = \left[
	%	\mathbf{0}_{jN_\mathrm{RF}\times N_\mathrm{RF}}^\textsf{T},
	%	\overline{\mathbf{W}}_j^\textsf{T},
	%	\mathbf{0}_{N-(j+1)N_\mathrm{RF}\times N_\mathrm{RF}}^\textsf{T} \right]^\textsf{T} \in \mathbb{C}^{N\times N_\mathrm{RF}}$, where $\overline{\mathbf{W}}_j\in \mathbb{C}^{N_\mathrm{RF}\times N_\mathrm{RF}}$ represents the combiner for the $j$-th block for $j = 1,\cdots, J$. Note that during collecting the received target echoes for $J = \frac{N}{N_\mathrm{RF}}$ time slots, the target DOAs are assumed to maintain invariant within a time slot while changing over time slots, which is reasonable for THz system wherein the symbol time in the order of picoseconds
	%Then, the $N_\mathrm{RF}\times T$ echo signal reflected from the $K$ targets at the $j$-th time slot is

	\subsection{Data Collection}
	To collect the full array data from $N_\mathrm{RF}$ RF chains, we follow a subarrayed approach. That is, the radar activates the antennas in a subarrayed fashion to obtain $N\times 1$ array data in $J=\frac{N}{N_\mathrm{RF}}$ time slots.  This is consistent with THz radar wherein the coherence  time is of the order of picoseconds~\cite{milliDegree_doa_THz_Chen2021Aug,mimoRadar_WidebandYu2019May}. Let $\mathbf{W}_j \in \mathbb{C}^{N\times N_\mathrm{RF}}$ be the applied combiner matrix at the $j$-th time slot (instead of $\breve{\mathbf{W}}$ in (\ref{radarReceived1})) as $\mathbf{W}_j = \left[
	\mathbf{0}_{jN_\mathrm{RF}\times N_\mathrm{RF}}^\textsf{T},
	\overline{\mathbf{W}}_j^\textsf{T},
	\mathbf{0}_{N-(j+1)N_\mathrm{RF}\times N_\mathrm{RF}}^\textsf{T} \right]^\textsf{T} \in \mathbb{C}^{N\times N_\mathrm{RF}}$, where $\overline{\mathbf{W}}_j\in \mathbb{C}^{N_\mathrm{RF}\times N_\mathrm{RF}}$ represents the combiner for the $j$-th block for $j = 1,\cdots, J$. Then, the $N_\mathrm{RF}\times T$ echo signal reflected from $K$ targets at the $j$-th time slot is
	\begin{align}
	\label{radarReceived}
	{\mathbf{Y}}_{j,m}  &= \sum_{k = 1}^{K} \mathbf{W}_j^\textsf{H} {\mathbf{a}} (\bar{\theta}_{m,k},\bar{r}_{m,k}) \widetilde{\mathbf{x}}_{m,k} + \mathbf{W}_j^\textsf{H}\widetilde{\mathbf{N}}_m \nonumber \\
	&	= 
	\sum_{k = 1}^K \beta_{m,k} \mathbf{W}_j^\textsf{H}{\mathbf{a}}_{m,k} {\mathbf{a}}_{m,k}^\textsf{T} {\mathbf{X}}_m + {\mathbf{N}}_{j,m},
	\end{align}
	where ${\mathbf{a}}_{m,k} = \mathbf{a}(\bar{\theta}_{m,k},\bar{r}_{m,k})\in \mathbb{C}^{N}$ and    ${\mathbf{N}}_{j,m} = \mathbf{W}_j^\textsf{H}\widetilde{\mathbf{N}}_m\in \mathbb{C}^{N_\mathrm{RF}\times T}$ represents the noise term. Define
	\begin{align}
	&\mathbf{A}_m = \left[\mathbf{a}_{m,1},\cdots, \mathbf{a}_{m,K}   \right]\in \mathbb{C}^{N\times K}, \\ 
	&{\mathbf{D}}_{j,m} = \mathbf{W}_j^\textsf{H}\mathbf{A}_m\in \mathbb{C}^{N_\mathrm{RF}\times K}, \\
	&\boldsymbol{\Pi}_m = \mathrm{diag}\{\beta_{m,1}, \cdots, \beta_{m,K} \}\in \mathbb{C}^{K\times K},
	\end{align}
	Then, (\ref{radarReceived}) becomes
	\begin{align}
	\label{arrayData}
	{\mathbf{Y}}_{j,m} = {\mathbf{D}}_{j,m}    \boldsymbol{\Pi}_m {\mathbf{A}}_m^\textsf{T}{\mathbf{X}}_m + {\mathbf{N}}_{j,m}.
	\end{align}
	Stacking all $\mathbf{Y}_{j,m}$ into a single $N\times T$ matrix leads to the overall observation matrix $\mathbf{Y}_m\in \mathbb{C}^{N\times T}$ as $	\mathbf{Y}_m = \left[ \mathbf{Y}_{1,m}^\textsf{T}, \cdots, \mathbf{Y}_{J,m}^\textsf{T} \right]^\textsf{T} $, i.e.,
	\begin{align}
	\mathbf{Y}_m= \mathbf{D}_m \boldsymbol{\Pi}_m \mathbf{A}_m^\textsf{T} \mathbf{X}_m + {\mathbf{N}}_m, \label{obs1}
	\end{align}
	where $\mathbf{D}_m = \left[\mathbf{D}_{1,m}^\textsf{T},\cdots, \mathbf{D}_{J,m}^\textsf{T} \right]^\textsf{T} = \mathbf{W}^\textsf{H}\mathbf{A}_m \in \mathbb{C}^{N\times K}$ $\mathbf{W} = \left[{\mathbf{W}}_1,\cdots, {\mathbf{W}}_J  \right]\in \mathbb{C}^{N\times N}$ and ${\mathbf{N}}_m = \left[{\mathbf{N}}_{1,m}^\textsf{T},\cdots, {\mathbf{N}}_{J,m}^\textsf{T}  \right]^\textsf{T}$. The $N\times T$ array output data in (\ref{obs1}) is collected via limited number of RF chains from  multiple time slots. This is used to construct the covariance matrix to invoke the MUSIC algorithm.

	%	, as it will be introduced in the following.

	\subsection{Parameter Estimation}
	
	Next, we  introduce the corrected noise subspace for beam-squint to accurately estimate the physical DoA and ranges. Define the ${N\times N}$ covariance matrix of the observations in (\ref{obs1}) as $\mathbf{R}_m= \frac{1}{T}{\mathbf{Y}}_m {\mathbf{Y}}_m^\textsf{H}$, i.e., 
	\begin{align}
	\mathbf{R}_m & = \frac{1}{T} {\mathbf{D}}_m \left( \frac{P_rT}{MN}\widetilde{\boldsymbol{\Pi} }_m\right) {\mathbf{D}}_m^\textsf{H} +  \frac{1}{T}{\mathbf{N}}_m{\mathbf{N}}_m^\textsf{H} \nonumber\\
	& \approxeq \frac{P_r}{MN} {\mathbf{D}}_m \widetilde{\boldsymbol{\Pi}}_m {\mathbf{D}}^\textsf{H}_m  +  \frac{\sigma^2}{N}  \mathbf{I}_{{N}}, \label{R_m1}
	\end{align}
	where ${\mathbf{N}}_m{\mathbf{N}}_m^\textsf{H} \approxeq \frac{{\sigma}^2 T}{N} \mathbf{I}_{N} $ (because $\mathbf{W}^\textsf{H}\mathbf{W} = \frac{1}{N}$) and  
	\begin{align}
	\widetilde{\boldsymbol{\Pi} }_m=  \boldsymbol{\Pi}_m\mathbf{A}_m^\textsf{T}\mathbf{A}_m^*\boldsymbol{\Pi}_m^\textsf{H} \in \mathbb{C}^{K\times K}.
	\end{align} The eigendecomposition of $\mathbf{R}_m$ yields $	\mathbf{R}_m = \mathbf{U}_m \boldsymbol{\Sigma}_m \mathbf{U}_m^\textsf{H},$
	where $\boldsymbol{\Sigma}_m\in \mathbb{C}^{N\times N}$ is a diagonal matrix composed of the eigenvalues of $\mathbf{R}_m$ in a descending order; $\mathbf{U}_m = \left[\mathbf{U}_{m}^\mathrm{S}\hspace{2pt} \mathbf{U}_m^\mathrm{N} \right]\in \mathbb{C}^{N\times N}$ corresponds to the eigenvector matrix; $\mathbf{U}_m^\mathrm{S}\in\mathbb{C}^{N\times K}$ and $\mathbf{U}_m^\mathrm{N}\in \mathbb{C}^{N\times N-K}$ are the signal and noise subspace eigenvector matrices, respectively. 
	
	By exploiting the orthogonality of the signal and noise subspaces, i.e.,  $	\mathbf{U}_m^\mathrm{N} \perp  \mathbf{U}_m^\mathrm{S}$, and the fact that  the columns of $ \mathbf{U}_m^\mathrm{S}$ and $\mathbf{D}_m$ span the same subspace~\cite{music_Schmidt1986Mar,friedlander}, we have
	\begin{align}
	\| \mathbf{d}_{m,k}^\textsf{H}{\mathbf{U}_m^\mathrm{N}}  \|_2^2=0, \label{perp1}
	\end{align}
	where  $\mathbf{d}_{m,k} = \mathbf{W}^\textsf{H} \mathbf{a}(\bar{\theta}_{m,k},
	\bar{r}_{m,k}) \in \mathbb{C}^{N} $ is the $k$-th column of $\mathbf{D}_m\in \mathbb{C}^{N\times K}$. 
	
	Note that (\ref{perp1}) implies the orthogonality with the corrupted steering vector $\mathbf{d}_{m,k}$, whereas our aim is to estimate the beam-squint-free physical DOA/ranges $\theta_k$, $r_k$. Therefore, define  ${\mathbf{V}_m^\mathrm{N}}\in \mathbb{C}^{N\times N-K}$ as the beam-squint-corrected noise subspace matrix, which is orthogonal to the nominal steering vectors. To that end, denote the beam-squint transformation matrix by $\mathbf{T}_{m}(\theta_{k},r_{k})\in \mathbb{C}^{N\times N}$. This provides a linear mapping between the nominal and beam-squint-corrupted steering vectors as 
	\begin{align}
	\mathbf{a}(\bar{\theta}_{m,k},\bar{r}_{m,k}) = 	\mathbf{T}_{m}(\theta_{k},r_{k})\mathbf{a}(\theta_k,r_k), \label{transformation}
	\end{align}
	where $\mathbf{T}_{m}(\theta_{k},r_{k}) = \mathrm{diag}\{\boldsymbol{\tau}_{m}(\theta_{k},r_{k})  \}$, 	for which the $n$-th element of $\boldsymbol{\tau}_{m}(\theta_{k},r_{k})\in \mathbb{C}^N$ is 
	$\tau_{m,k,n} = [\mathbf{a}(\bar{\theta}_{m,k},\bar{r}_{m,k})]_n/[\mathbf{a}(\theta_k,r_k)]_n$.
	Using (\ref{transformation}),  (\ref{perp1}) becomes
	\begin{align}
	&\| \left(  \mathbf{W}^\textsf{H} {\mathbf{a}}(\bar{\theta}_{m,k},\bar{r}_{m,k})  \right)^\textsf{H} {\mathbf{U}_m^\mathrm{N}}\|_2^2  \nonumber \\
	&=\|  \mathbf{a}^\textsf{H}({\theta}_{k},r_k) {\mathbf{T}_{m}^\textsf{H}(\theta_{k},r_{k})   \mathbf{W} {\mathbf{U}_m^\mathrm{N}}}\|_2^2 \nonumber \\
	&=\|  \mathbf{a}^\textsf{H} ({\theta}_{k},r_k)\mathbf{V}_m^\mathrm{N}     	 \|_2^2 = 0, \label{perp2}
	\end{align}
	where $	\mathbf{V}_m^\mathrm{N}$ is the corrected noise subspace matrix, i.e., 
	\begin{align}
	\mathbf{V}_m^\mathrm{N} \triangleq   \mathbf{T}_{m,k}^\textsf{H} \mathbf{G}_m^\textsf{H} \mathbf{W} {\mathbf{U}_m^\mathrm{N}}.
	\end{align}
	Examining (\ref{perp2}) reveals the useful property regarding the orthogonality of the corrected noise subspace $\mathbf{V}_m^\mathrm{N} $ and the  beam-squint-free steering vectors as $\mathbf{a}(\theta_{k},r_k) \perp \mathbf{V}_m^\mathrm{N} $ for $m\in \mathcal{M}$. Consequently,  we write the beam-squint-corrected MUSIC spectra for $M$ subcarriers as 
	\begin{align}
	\label{musicSpectra2}
	P(\theta,r) = \sum_{m = 1}^M  \frac{1}{\mathbf{a}^\textsf{H}(\theta,r)\mathbf{V}_m^\mathrm{N}{\mathbf{V}_m^\mathrm{N}}^\textsf{H} \mathbf{a}(\theta,r)},
	\end{align}
	whose $K$ highest peaks correspond to the physical target DoAs/ranges $\{\hat{{\theta}}_k, \hat{r}_k\}_{k=1}^K$, which can be identified through a peak-finding algorithm for (\ref{musicSpectra2}) only once because it includes the combination of spectra for $M$ subcarriers. 
		
	%-------------------------------------------------------------------------------------------------
	\begin{algorithm}[h]
		%				\footnotesize
		\begin{algorithmic}[1] 
			\caption{ \bf Near-field MUSIC }
			\Statex {\textbf{Input:}  $\mathbf{Y}_m$, $\mathbf{W}$, $K$, $\Psi$, $\mathcal{R}$, $\eta_m$ for $m\in \mathcal{M}$.} \label{alg}
			\Statex {\textbf{Output:} $\{\hat{\theta}_k,\hat{r}_k\}_{k =1}^{K}$.  } 
			\State  $\mathbf{R}_m= \frac{1}{T}{\mathbf{Y}}_m {\mathbf{Y}}_m^\textsf{H}$ for $m\in \mathcal{M}$.
			\State   Obtain the noise subspace $\mathbf{U}_m^\mathrm{N}$ from $\mathbf{R}_m$ for $m\in \mathcal{M}$. 
			\State  \textbf{for} $\theta \in \Psi$ and $r\in \mathcal{R}$ \textbf{do}
			\State \indent \textbf{for} $m\in\mathcal{M}$ 
			\State \indent Construct $\mathbf{a}(\theta,r)$ and $\mathbf{a}(\bar{\theta}_{m},\bar{r}_{m})$ from (\ref{steeringVectorPhy2}) and (\ref{steeringVectorSpa}).
			\State  \indent  Construct $\mathbf{T}_{m}(\theta,r)$ as in (\ref{transformation}).
			\State\indent  
			${\mathbf{V}_m^\mathrm{N}} \gets \mathbf{T}_m^\textsf{H}(\theta,r)   \mathbf{W} {\mathbf{U}_m^\mathrm{N}}$.
			\State\indent  $P_m(\theta,r) \gets \frac{1}{\mathbf{a}^\textsf{H}(\theta,r){\mathbf{V}_m^\mathrm{N}}{{\mathbf{V}_m^\mathrm{N}}}^\textsf{H} \mathbf{a}(\theta,r)}.$
			\State \indent \textbf{end for}  
			\State   \textbf{end for}
			\State  Combined MUSIC spectra: ${P}(\theta,r)  \gets \sum_{m = 1}^M {P}_m(\theta,r)$.
			\State  Find $\{\hat{\theta}_k,\hat{r}_k\}_{k =1}^{K}$ from the $K$ highest peaks of ${P}(\theta,r)$.
		\end{algorithmic} 
	\end{algorithm}
	%------------------------------------------------------------------------------------------------
	
	Algorithm~\ref{alg} presents the algorithmic steps for the proposed  approach. Specifically, we first compute the beam-squint-corrupted noise subspace $\mathbf{U}_m^\mathrm{N}$ and the beam-squint transformation matrix $\mathbf{T}_m(\theta,r)$ for $\theta\in \Psi = [-1,1]$ and $r\in \mathcal{R}=[0,d_F]$ in Steps 2-5. By constructing the corrected noise subspace  $	\mathbf{V}_m^\mathrm{N}$ in Step 7, the two-dimensional (2-D) MUSIC spectra $P_m(\theta,r)$ is computed in Step 8 for the $m$-th subcarrier. Then, the estimated DoA angles and ranges $\hat{\theta}_k,\hat{r}_k$ are computed from the combined 2-D MUSIC spectra $P(\theta,r)$ in Step 11.

		\begin{figure}[t]
		\centering
		{\includegraphics[draft=false,width=\columnwidth]{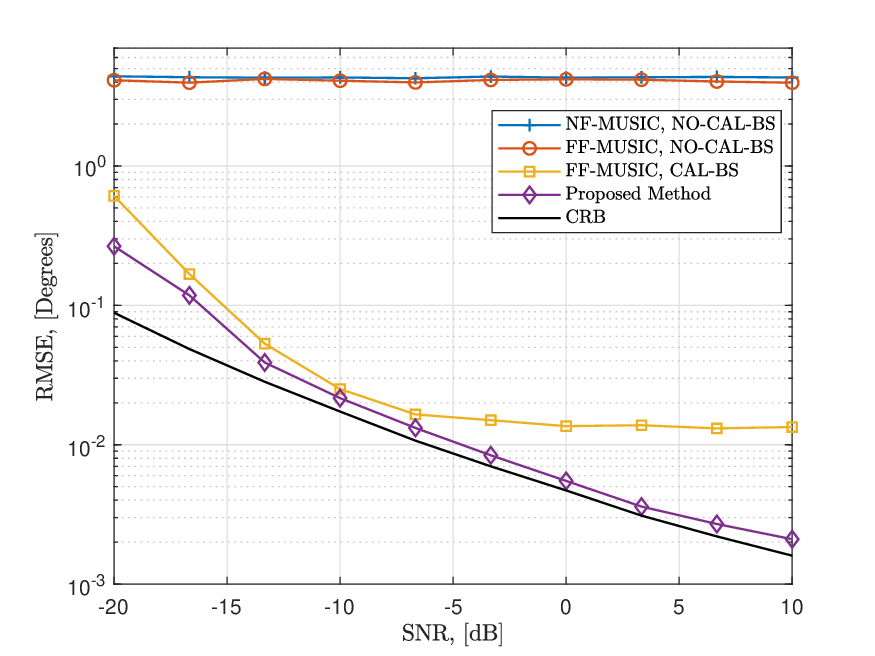} } 
		\caption{ DoA estimation RMSE vs. SNR. 
		}
		%			\vspace*{-5mm}
		\label{fig_DOA_RMSE_SNR}
	\end{figure}
	%%-----------------------------------------------------	
	
	\subsection{Computational Complexity and Identifiability}
	%	\subsubsection{Computational Complexity} 
	The complexity of the proposed approach is mainly due to eigendecomposition of $\mathbf{R}_m$ ($O(MN^3)$) as well as the computation the corrected noise subspace $\mathbf{V}_m^\mathrm{N}$ ($O(MN^2[3N -K])$) for $m\in \mathcal{M}$. Thus, the overall computational complexity order is $O(MN^2[4N-K])$. Note that the complexity reduces to  $O(2MN^3)$ for the traditional MUSIC algorithm, which does not account for beam-squint. The  problem of target localization involves $K$ coupled (DoA and range) unknowns, while the collected array data from $N_\mathrm{RF}$ RF chains for $J = \frac{N}{N_\mathrm{RF}}$ time-slots is $N\times 1$ for $M$ subcarriers.  Hence, the proposed technique  is feasible only if $\mathrm{rank} \{\mathbf{U}_m^\mathrm{N}{\mathbf{U}_m^\mathrm{N}}^\textsf{H} \} = N - K \geq 1$, provided that $T \geq K$ data snapshots are available. This condition becomes $N_\mathrm{RF}-K \geq 1$ if the  output for a single time-slot is used.

	\section{Numerical Experiments}
	The efficiency of our proposed method is benchmarked against the far-field and near-field MUSIC algorithms and beam-squint-corrected MUSIC algorithm as well as the Cram\'er-Rao bound (CRB)~\cite{elbir_THZ_CE_ArrayPerturbation_Elbir2022Aug} in terms of  root-MSE (RMSE), i.e., $\mathrm{RMSE}_\theta = (\frac{1}{J_TK} \sum_{i=1}^{J_T}\sum_{k=1}^K | \hat{{\theta}}_{i,k}- {{\theta}}_{i,k}|^2 )^{1/2}$ and $\mathrm{RMSE}_r = (\frac{1}{J_TK} \sum_{i=1}^{J_T}\sum_{k=1}^K | \hat{{r}}_{i,k}- {{r}}_{i,k}|^2 )^{1/2}$, where $\hat{{\theta}}_{i,k}$ and $\hat{r}_{i,k}$ stand for the estimated DoA and the ranges, respectively, for the $i$-th instance of $J_T= 500$ Monte Carlo trials. The default simulation parameters are $f_c=300$ GHz, $B=30$ GHz, $M=32$, $N = 128$, $N_\mathrm{RF}=8$, $T=500$, $K=2$~\cite{milliDegree_doa_THz_Chen2021Aug,delayPhasePrecoding_THz_Dai2022Mar}. The DoAs are selected uniform at random from the interval $\tilde{\theta}_{k}\sim \mathrm{unif} [-\frac{\pi}{2},\frac{\pi}{2}]$. The combiner matrix is modeled as $[\mathbf{W}]_{i,j} = \frac{1}{\sqrt{N}}e^{\mathrm{j}{\Phi}}$, where ${\Phi} \sim \text{unif}[-\frac{\pi}{2},\frac{\pi}{2}]$ for $i \in [1, N]$ and $j  \in [1,  N_\mathrm{RF}]$.

	Fig.~\ref{fig_DOA_RMSE_SNR} shows the DoA estimation RMSE with respect to signal-to-noise ratio (SNR), defined as $\mathrm{SNR} = 10\log_{10}(\frac{\rho}{\sigma_n^2})$ with $\rho = \frac{P_{r}}{M^2N^2} = 1$. We see that both near-field (NF) and far-field (FF) MUSIC algorithms display poor performance ($\sim 4^\circ$) without beam-squint calibration (NO-CAL-BS) whereas FF-MUSIC achieves relatively lower DoA estimation RMSE ($\sim 0.015^\circ$) when beam-squint is perfectly calibrated (CAL-BS). This observation shows that the impact of beam-squint is more severe than the near/far-field model mismatch. While the a priori calibration of beam-squint significantly reduces the DoA estimation RMSE, the proposed approach outperforms the competing algorithms by attaining the CRB very closely.  This superior performance of the proposed approach can be attributed to the calibration of near-field beam-squint without any priori knowledge, thereby proving high resolution DoA accuracy.

%	
%	Fig.~\ref{fig_DOA_RMSE_B}  shows the DoA estimation RMSE versus bandwidth $B$. In comparison, beam-squint-corrupted and beam-squint-calibrated MUSIC algorithms lead to approximately $4^\circ$ and $0.015^\circ$ DoA error while the proposed method provides accurate DoA estimation RMSE for wide range of bandwidth, i.e., $B\in [0,30]$ GHz. 
%	
%	
%	
%	
%	%%-----------------------------------------------------
%	\begin{figure}[t]
%		\centering
%		{\includegraphics[draft=false,width=\columnwidth]{RMSE_B.eps} } 
%		\caption{ DoA estimation RMSE vs. bandwidth $B$ for  $\mathrm{SNR} = 0$ dB.
%			\vspace*{-3.5cm}}
%		%			\vspace*{-5mm}
%		\label{fig_DOA_RMSE_B}
%	\end{figure}
%	%%-----------------------------------------------------	
%	

	\section{Summary}
	We investigated the near-field DoA estimation problem for THz automotive radar. We propose a MUSIC-like approach with noise subspace correction in order to compensate for the beam-squint. The performance of the proposed method is evaluated in the presence of near-field beam-squint. It is shown that the DoA error due to beam-squint is more severe ($\sim 4^\circ$) than that of near/far model mismatch ($\sim 0.01^\circ$). In contrast, our proposed method exhibits accurate DoA estimation performance with high precision.

	%	\newpage
	\balance
	\bibliographystyle{IEEEtran}
	\bibliography{references_131,references_121}

\begin{thebibliography}{10}
\providecommand{\url}[1]{#1}
\csname url@samestyle\endcsname
\providecommand{\newblock}{\relax}
\providecommand{\bibinfo}[2]{#2}
\providecommand{\BIBentrySTDinterwordspacing}{\spaceskip=0pt\relax}
\providecommand{\BIBentryALTinterwordstretchfactor}{4}
\providecommand{\BIBentryALTinterwordspacing}{\spaceskip=\fontdimen2\font plus
\BIBentryALTinterwordstretchfactor\fontdimen3\font minus
  \fontdimen4\font\relax}
\providecommand{\BIBforeignlanguage}[2]{{%
\expandafter\ifx\csname l@#1\endcsname\relax
\typeout{** WARNING: IEEEtran.bst: No hyphenation pattern has been}%
\typeout{** loaded for the language `#1'. Using the pattern for}%
\typeout{** the default language instead.}%
\else
\language=\csname l@#1\endcsname
\fi
#2}}
\providecommand{\BIBdecl}{\relax}
\BIBdecl

\bibitem{thz_Akyildiz2022May}
I.~F. Akyildiz, C.~Han, Z.~Hu, S.~Nie, and J.~M. Jornet, ``{Terahertz Band
  Communication: An Old Problem Revisited and Research Directions for the Next
  Decade},'' \emph{IEEE Trans. Commun.}, vol.~70, no.~6, pp. 4250--4285, May
  2022.

\bibitem{milliDegree_doa_THz_Chen2021Aug}
Y.~Chen, L.~Yan, C.~Han, and M.~Tao, ``{Millidegree-Level Direction-of-Arrival
  Estimation and Tracking for Terahertz Ultra-Massive MIMO Systems},''
  \emph{IEEE Trans. Wireless Commun.}, vol.~21, no.~2, pp. 869--883, Aug. 2021.

\bibitem{milliDegreeDOA_THz_Peng2016Aug}
B.~Peng and T.~K{\ifmmode\ddot{u}\else\"{u}\fi}rner, ``{Three-Dimensional Angle
  of Arrival Estimation in Dynamic Indoor Terahertz Channels Using a
  Forward{\textendash}Backward Algorithm},'' \emph{IEEE Trans. Veh. Technol.},
  vol.~66, no.~5, pp. 3798--3811, Aug. 2016.

\bibitem{elbir2022Aug_THz_ISAC}
A.~M. Elbir, K.~V. Mishra, S.~Chatzinotas, and M.~Bennis, ``{Terahertz-Band
  Integrated Sensing and Communications: Challenges and Opportunities},''
  \emph{arXiv}, Aug. 2022.

\bibitem{thzRadar1_Bhattacharjee}
S.~Bhattacharjee, K.~V. Mishra, R.~Annavajjala, and C.~R. Murthy,
  ``{Multi-Carrier Wideband OCDM-Based THZ Automotive Radar},'' in
  \emph{{ICASSP 2023 - 2023 IEEE International Conference on Acoustics, Speech
  and Signal Processing (ICASSP)}}.\hskip 1em plus 0.5em minus 0.4em\relax
  IEEE, pp. 04--10.

\bibitem{thzRadar2_Xiao2020Mar}
Y.~Xiao, F.~Norouzian, E.~G. Hoare, E.~Marchetti, M.~Gashinova, and
  M.~Cherniakov, ``{Modeling and Experiment Verification of Transmissivity of
  Low-THz Radar Signal Through Vehicle Infrastructure},'' \emph{IEEE Sens. J.},
  vol.~20, no.~15, pp. 8483--8496, Mar. 2020.

\bibitem{elbir2021JointRadarComm}
A.~M. Elbir, K.~V. Mishra, and S.~Chatzinotas, ``{Terahertz-Band Joint
  Ultra-Massive MIMO Radar-Communications: Model-Based and Model-Free Hybrid
  Beamforming},'' \emph{IEEE J. Sel. Top. Signal Process.}, vol.~15, no.~6, pp.
  1468--1483, Oct. 2021.

\bibitem{eamaz2023near}
A.~Eamaz, F.~Yeganegi, K.~V. Mishra, and M.~Soltanalian, ``Near-field
  low-{WISL} unimodular waveform design for terahertz automotive radar,'' in
  \emph{European Signal Processing Conference}, 2023, pp. 1--5.

\bibitem{mishra2023signal}
K.~V. Mishra, I.~Bilik, J.~Tabrikian, and A.~P. Petropulu, ``Signal processing
  for terahertz-band automotive radars: {E}xploring the next frontier,''
  \emph{arXiv preprint}, 2023.

\bibitem{ummimoTareqOverview}
H.~Sarieddeen, M.-S. Alouini, and T.~Y. Al-Naffouri, ``An overview of signal
  processing techniques for {T}erahertz communications,'' \emph{Proceedings of
  the IEEE}, vol. 109, no.~10, pp. 1628--1665, 2021.

\bibitem{elbir2022Nov_SPM_beamforming}
A.~M. Elbir, K.~V. Mishra, S.~A. Vorobyov, and R.~W. Heath, ``{Twenty-Five
  Years of Advances in Beamforming: From convex and nonconvex optimization to
  learning techniques},'' \emph{IEEE Signal Process. Mag.}, vol.~40, no.~4, pp.
  118--131, Jun. 2023.

\bibitem{delayPhasePrecoding_THz_Dai2022Mar}
L.~Dai, J.~Tan, Z.~Chen, and H.~V. Poor, ``{Delay-Phase Precoding for Wideband
  THz Massive MIMO},'' \emph{IEEE Trans. Wireless Commun.}, p.~1, Mar. 2022.

\bibitem{beamSquint_FeiFei_Wang2019Oct}
B.~Wang, M.~Jian, F.~Gao, G.~Y. Li, and H.~Lin, ``{Beam Squint and Channel
  Estimation for Wideband mmWave Massive MIMO-OFDM Systems},'' \emph{IEEE
  Trans. Signal Process.}, vol.~67, no.~23, pp. 5893--5908, Oct. 2019.

\bibitem{elbir_THZ_CE_ArrayPerturbation_Elbir2022Aug}
A.~M. Elbir, W.~Shi, A.~K. Papazafeiropoulos, P.~Kourtessis, and
  S.~Chatzinotas, ``{Terahertz-Band Channel and Beam Split Estimation via Array
  Perturbation Model},'' \emph{IEEE Open J. Commun. Soc.}, p.~1, Mar. 2023.

\bibitem{thz_beamSplit}
J.~Tan and L.~Dai, ``{Wideband Beam Tracking in THz Massive MIMO Systems},''
  \emph{IEEE J. Sel. Areas Commun.}, vol.~39, no.~6, pp. 1693--1710, Apr 2021.

\bibitem{calibrationMassiveMIMOWei2020Apr}
X.~Wei, Y.~Jiang, Q.~Liu, and X.~Wang, ``{Calibration of Phase Shifter Network
  for Hybrid Beamforming in mmWave Massive MIMO Systems},'' \emph{IEEE Trans.
  Signal Process.}, vol.~68, pp. 2302--2315, Apr. 2020.

\bibitem{doaEst_mmWave_Zhang2021Oct}
R.~Zhang, B.~Shim, and W.~Wu, ``{Direction-of-Arrival Estimation for Large
  Antenna Arrays With Hybrid Analog and Digital Architectures},'' \emph{IEEE
  Trans. Signal Process.}, vol.~70, pp. 72--88, Oct. 2021.

\bibitem{norouzian2019rain}
F.~Norouzian, E.~Marchetti, M.~Gashinova, E.~Hoare, C.~Constantinou,
  P.~Gardner, and M.~Cherniakov, ``Rain attenuation at millimeter wave and
  low-{THz} frequencies,'' vol.~68, no.~1, pp. 421--431, 2019.

\bibitem{nf_primer_Bjornson2021Oct}
E.~Bj{\ifmmode\ddot{o}\else\"{o}\fi}rnson, {\ifmmode\ddot{O}\else\"{O}\fi}.~T.
  Demir, and L.~Sanguinetti, ``A primer on near-field beamforming for arrays
  and reconfigurable intelligent surfaces,'' in \emph{Asilomar Conference on
  Signals, Systems, and Computers}, Oct. 2021, pp. 105--112.

\bibitem{elbir2023Feb_NF_THZ_CE_ICASSP_NBAOMP}
A.~M. Elbir, K.~V. Mishra, and S.~Chatzinotas, ``{NBA-OMP: Near-field
  Beam-Split-Aware Orthogonal Matching Pursuit for Wideband THz Channel
  Estimation},'' \emph{arXiv}, Feb. 2023.

\bibitem{nf_OMP_Dai_Wei2021Nov}
X.~Wei and L.~Dai, ``Channel estimation for extremely large-scale massive
  {MIMO}: {F}ar-field, near-field, or hybrid-field?'' \emph{IEEE Commun.
  Lett.}, vol.~26, no.~1, pp. 177--181, Nov. 2021.

\bibitem{nf_mmwave_CE_noBeamSplit_Cui2022Jan}
M.~Cui and L.~Dai, ``Channel estimation for extremely large-scale {MIMO}:
  {F}ar-field or near-field?'' \emph{IEEE Trans. Commun.}, vol.~70, no.~4, pp.
  2663--2677, Jan. 2022.

\bibitem{nf_NB2_Zhang2022Nov}
X.~Zhang, Z.~Wang, H.~Zhang, and L.~Yang, ``Near-field channel estimation for
  extremely large-scale array communications: {A} model-based deep learning
  approach,'' \emph{IEEE Communications Letters}, vol.~27, no.~4, pp.
  1155--1159, 2023.

\bibitem{lv2022co}
W.~Lv, K.~V. Mishra, and S.~Chen, ``Co-pulsing {FDA} radar,'' \emph{IEEE
  Transactions on Aerospace and Electronic Systems}, 2022, in press.

\bibitem{lv2022clutter}
------, ``Clutter suppression via space-time-range processing in co-pulsing
  {FDA} radar,'' in \emph{Asilomar Conference on Signals, Systems, and
  Computers}, 2022, pp. 470--475.

\bibitem{vouras2023overview}
P.~Vouras, K.~V. Mishra, A.~Artusio-Glimpse, S.~Pinilla, A.~Xenaki, D.~W.
  Griffith, and K.~Egiazarian, ``An overview of advances in signal processing
  techniques for classical and quantum wideband synthetic apertures,''
  \emph{IEEE Journal of Selected Topics in Signal Processing}, vol.~17, no.~2,
  pp. 317--369, 2023.

\bibitem{vouras2023phase}
P.~Vouras, K.~V. Mishra, and A.~Artusio-Glimpse, ``Phase retrieval for
  {R}ydberg quantum arrays,'' in \emph{IEEE International Conference on
  Acoustics, Speech and Signal Processing}, 2023, pp. 1--5.

\bibitem{music_Schmidt1986Mar}
R.~Schmidt, ``{Multiple emitter location and signal parameter estimation},''
  \emph{IEEE Trans. Antennas Propag.}, vol.~34, no.~3, pp. 276--280, Mar. 1986.

\bibitem{mimoRadar_WidebandYu2019May}
X.~Yu, G.~Cui, J.~Yang, L.~Kong, and J.~Li, ``Wideband {MIMO} radar waveform
  design,'' \emph{IEEE Trans. Signal Process.}, vol.~67, no.~13, pp.
  3487--3501, 2019.

\bibitem{nf_Fresnel_Cui2022Nov}
M.~Cui, L.~Dai, Z.~Wang, S.~Zhou, and N.~Ge, ``Near-field rainbow: {W}ideband
  beam training for {XL-MIMO},'' \emph{IEEE Transactions on Wireless
  Communications}, vol.~22, no.~6, pp. 3899--3912, Nov. 2023.

\bibitem{elbir2023Feb_NF_THZ_CE}
A.~M. Elbir, W.~Shi, A.~K. Papazafeiropoulos, P.~Kourtessis, and
  S.~Chatzinotas, ``Near-field terahertz communications: {M}odel-based and
  model-free channel estimation,'' \emph{IEEE Access}, vol.~11, pp.
  36\,409--36\,420, Apr. 2023.

\bibitem{friedlander}
B.~Friedlander and A.~J. Weiss, ``{Direction finding in the presence of mutual
  coupling},'' \emph{IEEE Trans. Antennas Propag.}, vol.~39, no.~3, pp.
  273--284, Mar. 1991.

\end{thebibliography}

\end{document}